\documentstyle[12pt]{article}

\begin{document}
\newcommand{\hs}{\hspace{5mm}}
\newcommand{\vs}{\vspace{2mm}}
\newcommand{\alphap}{{\alpha}^{\prime}}
\newcommand{\nup}{{\nu}^{\prime}}
\newcommand{\alphaz}{{\alpha}^{\prime \prime}}
\newcommand{\nuz}{{\nu}^{\prime \prime}}
\newcommand{\z}{\prime \prime}
\newcommand{\p}{\prime}
\begin{titlepage}
\vspace{-10mm}
\begin{flushright}

\end{flushright}
\vspace{12pt}
\begin{center}
\begin{large}
{\bf On the gravitational coupling of Hadamard states  }\\
\end{large}
\vspace{5mm} ${\bf Hadi~Salehi}, {\bf P.~ Moyassari
}\footnote{e-mail:~p-moyassari@cc.sbu.ac.ir},\vspace{1cm} {\bf R.~
Rashidi}$

 Department of Physics, Shahid Beheshti university \\
Tehran 19839,  IRAN. \\  \vspace{0.3cm}
\end{center}
\abstract{We study the constraints imposed by the Hadamard condition
on the two-point function of local states of a scalar quantum field
conformally coupled to a gravitational background. We propose a
method to assign a stress tensor to the state-dependent part of the
two point function which arises as a conserved tensor with an
anomalous trace. To characterize the local Hadamard states of
physical interest we apply a super-selection rule relating this
quantum stress tensor to the matter stress tensor of a conformal
invariant gravitational model subject to a conformal symmetry
breaking term. This implies that the determination of a Hadamard
state may be considered as an integral part of its gravitational
coupling via the back-reaction effect. }
\end{titlepage}

\section{Introduction}
A central question in quantum field theory in curved space concerns
the determination of the physical states of quantized matter fields
propagating on a gravitational background. Usually one can assume
that the two-point function of the physical states have a local
structure corresponding to the Hadamard expansion \cite{1}. A state
characterized by this condition, a Hadamard state, is distinguished
by the fact that the singular part of the two-point function can
uniquely be determined by local geometry, but the regular part of
the two-point function does not admit a geometric construction and
must be considered as state-dependent. It is therefore essential for
the determination  of local states to apply super-selection rules
characterizing the regular part of the Hadamard states. Little is
known in the literature how this can be done in a physical
consistent manner. The present paper deals with this issue. As a
model we consider a quantum field conformally coupled to a
gravitational background. Taking the local states of this field as
Hadamard states we analyze  the constraints imposed on the regular
part (state-dependent part) of the two-point function. These
constraints are then used to assign a conserved quantum stress
tensor to a local state with an anomalous trace. To characterize the
local Hadamard states of physical interest we apply a
super-selection rule relating this quantum stress tensor to the
matter stress tensor of a conformal invariant gravitational model
subject to a conformal invariance breaking using a constant
mass-scale. This super-selection rule arises as a statement about
the nature of the gravitational coupling of a physical Hadamard
state, in that it requires an indispensable change of the underlying
background metric due to the back reaction of a Hadamard state. This
change is taken into account through a nontrivial conformal factor.
This means that a physical Hadamard state can be supported on a
physical metric which is conformally related to the underlying
background metric. The significance of this feature for the
asymptotic particle creation is demonstrated.

\section{ Hadamard state condition}

In this section we present the Hadamard prescription and briefly review the derivation of
 the local constraints on the state-dependent part of
 the two-point function of local states
 of a linear scalar quantum field $\phi$ conformally coupled to gravity
 with the action functional
 \begin{equation}
S[\phi]=-\frac{1}{2} \int d^{4}x g^{1/2}(g_{\alpha\beta}
\nabla^{\alpha}\phi \nabla^{\beta}\phi+\frac{1}{6} R \phi^{2}),
\label{1}
\end{equation}
where $R$ is the scaler curvature (In the following the semicolon
 and $\nabla$ indicate covariant differentiation ). This gives
 rise to the field equation:
 \begin{equation}
(\Box-\frac{1}{6} R)\phi(x)=0.  \label{2}
\end{equation}
A state of $\phi(x)$ is characterized by a hierarchy of
Wightman-functions (n-point functions)
\begin{equation}
 <\phi(x_1),\ldots,\phi(x_n)>. \label{a1}
\end{equation}
We are primarily interested in those states which reflect the
intuitive notion of a vacuum. For this aim, we may restrict
ourselves basically to quasi-free states for which the truncated
n-point functions vanish for $n>2$. Such states may be characterized
by their two-point functions. In a linear theory the anti-symmetric
part of the two-point function (commutator function) is common to
all states in the same representation. Therefore the individual
characteristics of a state are encoded in the symmetric part of the
two-point function denoted by $G^+(x,x')$. In Hadamard prescription
one assumes that $G^+(x,x')$ has a singular structure represented by
Hadamard expansion. This means that in a normal neighborhood of a
point $x$ the function $G^+(x,x')$ can be written as \cite{1,2}:
 \begin{equation}
G^{+}(x,x')=\frac{1}{8 \pi^{2}}
\{\frac{\Delta^{1/2}(x,x')}{\sigma(x,x')}+
V(x,x')~\ln\sigma(x,x')+W(x,x')\},  \label{3}
\end{equation}
 where $\sigma(x,x')$ is the square of the distance along
 the geodesic joining $x$ and $x'$ and
 \begin{equation}
\begin{array}{llll}\vspace{0.5cm}
\Delta(x,x')=-g^{-1/2}(x) Det\{-\sigma_{;\mu\nu }\}g^{-1/2}(x')\\
\vspace{0.5cm} g(x)=Det~ g_{\alpha\beta}\\\vspace{0.5cm}
V(x,x')=\sum_{n=0}^{+\infty} V_{n}(x,x') \sigma^n
\\\vspace{0.5cm} W(x,x')=\sum_{n=0}^{+\infty} W_{n}(x,x')
\sigma^n.  \label{4}
\end{array}
\end{equation}

 Applying (\ref{2}) to $G^+ (x,x')$ leads to some recursion
 relations in terms of $V_{n}$ and $W_{n}$. From these relations
 one can determine the function $V(x,x')$ uniquely in terms of
local geometry, but the function $W_0 (x,x')$ remains arbitrary
and its specification depends significantly on the choice of a
state \cite{3}. However, there is a general constraint on $W_0
(x,x')$ which can be obtained from the symmetry condition of $G^+
(x,x')$ and equation (\ref{2}). To get this constraint the
covariant expansion of $W_0 (x,x')$ is used only up to the third
order in $\sigma_{;\alpha}$ \cite{4}. In general there are some
additional constraints on the higher order expansion terms that in
our analysis are neglected. It was shown in details in \cite{4}
(also see \cite{3}) that the constraints imposed on the
state-dependent part of two-point function have the form
\begin{equation}
\Sigma_{\alpha\beta}^{;\alpha}=0, \label{5}
\end{equation}
where
\begin{equation}
\begin{array}{ll}\vspace{0.6cm}
\Sigma_{\alpha\beta}=(W_{0\alpha\beta}(x)-\frac{1}{4}g_{\alpha\beta}
W_{0\gamma}^{\gamma}(x))-\frac{1}{6}(R_{\alpha\beta}-\frac{1}{4}Rg_{\alpha\beta})W_{0}(x)\\
-\frac{1}{3}(W_{0;\beta\alpha}(x) -\frac{1}{4}g_{\alpha\beta} \Box
W_{0}(x))-\frac{1}{2}g_{\alpha\beta}v_1(x), \label{6}
\end{array}
\end{equation}
and
\begin{equation}
v_{1}(x)=\lim_{x'\rightarrow x}V_{1}(x,x')=\frac{1}{720}\{ \Box
R-R_{\alpha\beta}
R^{\alpha\beta}+R_{\alpha\beta\delta\gamma}R^{\alpha\beta\delta\gamma}\},\label{7}
\end{equation}
Functions $W_0$ and $W_{0\alpha\beta}$ are the coefficients in the
covariant expansion of $W_0 (x,x')$, we have namely
\begin{equation}
W_0(x,x')=W_0(x)-\frac{1}{2}W_{0;\alpha}(x)\sigma^{;\alpha}+\frac{1}{2}W_{0\alpha\beta}
(x)\sigma^{;\alpha}\sigma^{;\beta}+\textit{O}(\sigma^{3/2}).
\end{equation}
 It is obvious
from (\ref{6}) that
\begin {equation}
\Sigma^ \alpha_\alpha=-2v_1(x). \label{8}
\end {equation}
Since tensor $\Sigma_{\alpha\beta}$ arises as a conserved tensor
with an anomalous trace, we take
  it as the quantum stress tensor induced
by the two-point function \cite{5,6,7}. The determination of
$\Sigma_{\alpha\beta}$ is very essential for the characterization of
a local Hadamard state. Any assumption about $\Sigma_{\alpha\beta}$
which respects the constraints (\ref{5}) and (\ref{8}) acts as a
super-selection rule selecting a local Hadamard state and the
corresponding Hilbert space. The application of such a
super-selection rule has to respect the fundamental idea of general
relativity concerning the gravitational coupling of
$\Sigma_{\alpha\beta}$ . In order to specify the configuration of
$\Sigma_{\alpha\beta}$ we must therefore look at the equations
describing the gravitational coupling of $\Sigma_{\alpha\beta}$.

\section{The gravitational coupling}

For the characterization of the quantum stress tensor
$\Sigma_{\alpha\beta}$ we proceed to apply a super-selection rule in
form of a dynamical model that is taken to describe the
gravitational coupling of $\Sigma_{\alpha\beta}$ to a dynamical
metric $\bar{g}_{\alpha\beta}$. This super-selection rule relates
the quantum stress tensor $\Sigma_{\alpha\beta}$ to the matter
stress tensor of a conformal invariant gravitational model. To
arrive at the supers-election rule in question we first note that
the appearance of the anomalous trace in (\ref{8}) suggests that the
model should be defined in terms of a conformal invariance breaking.
We begin with the consideration of the action functional of a scalar
tensor theory, namely
  \begin{equation}
  S=-\frac{1}{2}\int
  d^4x\sqrt{-\bar{g}}(\bar{g}^{\alpha\beta}
  \bar{\nabla}_{\alpha}\psi\bar{\nabla}_{\beta}\psi+\frac{1}{6}\bar{R}\psi^{2}+\mu^2\psi^2)+S_{m}[\bar{g}^{\alpha\beta}].
  \label{9}
  \end{equation}
  in which $S_{m}$ stands for a matter action,
  $\bar{g}_{\alpha\beta}$ is a dynamical metric and $\bar{\nabla}$ is covariant
   derivative with respect to $\bar{g}_{\alpha\beta}$.
The parameter $\mu$ is a constant mass scale which implies the
conformal symmetry breaking through the term $\mu^{2}\psi^{2}$.
    Varying $S$ with
  respect to $\bar{g}^{\alpha\beta}$ and $\psi$ yields respectively:
  \begin{equation}
    \bar{G}_{\alpha\beta}-3\mu^2\bar{g}^{\alpha\beta}=-
    6\psi^{-2}(T_{\alpha\beta}+
    \tau_{\alpha\beta}),
    \label{10}
  \end{equation}
  \begin{equation}
  \bar{\Box}\psi-\frac{1}{6}\bar{R}\psi-\mu^2\psi=0,
  \label{11}
  \end{equation}
where
   \begin{equation}
  T_{\alpha\beta}=\frac{-2}{\sqrt{-\bar{g}}}\frac{\delta}{\delta
  \bar{g}^{\alpha\beta}}S_{m}[
\bar{g}^{\alpha\beta}],\label{13}
\end{equation}
 and
\begin{equation}
\tau_{\alpha\beta}=\bar{\nabla}_{\alpha}\psi\bar{\nabla}_{\beta}\psi-
\frac{1}{2}\bar{g}_{\alpha\beta}\bar{\nabla}_{\rho}\psi\bar{\nabla}^{\rho}\psi
+\frac{1}{6}(\bar{g}_{\alpha\beta}\bar{\Box}-\bar{\nabla}_{\alpha}\bar{\nabla}_{\beta})\psi^{2},\label{14}
\end{equation}
here $\bar{\Box}\equiv
\bar{g}^{\alpha\beta}\bar{\nabla}_{\alpha}\bar{\nabla}_{\beta}$ and
$\bar{G}_{\alpha\beta}$ is the Einstein tensor of metric
$\bar{g}_{\alpha\beta}$. Comparing the trace of (\ref{10}) with the
equation (\ref{11})
 yields,
\begin{equation}
\bar{g}^{\alpha\beta}T_{\alpha\beta}=
 \mu^2\psi^2. \label{17}
\end{equation}
Taking the four-divergence of (\ref{10}) and using (\ref{11}) leads
to
\begin{equation}
\bar{\nabla}^{\alpha} T_{\alpha\beta}=0. \label{16}
\end{equation}
It should be noted that in the derivation of the dynamical equations
no variation is performed with respect to the dynamical variables of
the matter action. In theories of this type the configuration of
matter is dynamically determined by the gravitational equations
(\ref{10}) once the metric is assumed to be fixed. We shall use this
feature to apply a super-selection rule relating the (still unknown)
quantum stress tensor $\Sigma_{\alpha\beta}$ to the matter stress
tensor $T_{\alpha\beta}$.

In accordance with this strategy we consider the equation (\ref{10})
as restricting the configuration of $T_{\alpha\beta}$, that is we
fix the metric $\bar{g}_{\alpha\beta}$ and interpret (\ref{10}) as
constraints on $T_{\alpha\beta}$\footnote{This corresponds to the
g-method \cite{7a} in interpreting the gravitational equations.}. We
then apply a super-selection rule of the form
\begin{equation}
\Sigma_{\alpha\beta}=T_{\alpha\beta}[\bar{g}_{\alpha\beta}].
\label{a4}
\end{equation}
This super-selection rule determind the gravitational coupling of
$\Sigma_{\alpha\beta}$ as
  \begin{equation}
 \Sigma_{\alpha\beta}= -(\frac{1}{6}\psi^{2} \bar{G}_{\alpha\beta}
   -3\mu^2\psi^{2}\bar{g}_{\alpha\beta}+ \tau_{\alpha\beta}),\label{a5}
   \end{equation}
which follows from the gravitational equation (\ref{10}) together
with (\ref{a4}).

The super-selection rule (\ref{a4}) is complete if we know the
metric $\bar{g}_{\alpha\beta}$. At this stage one may follow two
different methods. Firstly one may assume the metric
$\bar{g}_{\alpha\beta}$ to be equal to the background metric
$g_{\alpha\beta}$. If we follow this reasoning, then the constraint
(\ref{8}) may not be satisfied without recourse to the higher order
gravity constraints arising from the trace anomaly \cite{4}. In the
second method, according to the back reaction effect one may assume
that the gravitational coupling of a Hadamard state changes the
background metric. It means that a Hadamard state may not be
supported on a underlying background metric. In specific terms we
assume that this difference can be factorized in a conformal factor,
namely
\begin{equation}
\bar{g}_{\alpha\beta}=e^{-2\omega}g_{\alpha\beta}.\label{r}
\end{equation}

in which $\omega$ is a non-trivial conformal factor. Physically the
conformal factor $\omega$ may describe a change of the background
metric through a nontrivial back reaction effect of
 $\Sigma_{\alpha\beta}$. Therefore this approach emphasizes the
 indispensable role played by the back reaction in the
 characterization of Hadamard states. We shall follow
this reasoning in order not to be confronted with higher order
gravity constraints.

  The super-selection rule (\ref{a4}) should satisfy
the constraints (\ref{5}) and (\ref{8}) on the metric
$\bar{g}_{\alpha\beta}$. Because of (\ref{16}) the constraint
(\ref{5}) is automatically satisfied. To satisfy the other
constraint (\ref{8}) we get the consistency relation
\begin{equation}\label{b1}
    -2\bar{v}_1(x)=\mu^2\psi^2.
\end{equation}
In the four-dimensional case, the trace anomaly (\ref{7}) for the
metrics $g_{\alpha\beta}$ and $\bar{g}_{\alpha\beta}$ are related by
\cite{4}
\begin{equation}\label{b2}
\begin{array}{c}\vspace{0.5cm}
 \bar{v}_1(x)=\frac{1}{720}e^{4\omega}\{720v_1(x)+2R\Box\omega+2R_{;\alpha}\omega^{;\alpha}+
    6\Box(\Box\omega)+8(\Box\omega)^2 \\
    -8\omega_{;\alpha\beta}\omega^{;\alpha\beta} -8R_{;\alpha\beta}\omega^{;\alpha}\omega^{;\beta}-8\omega_{;\alpha}\omega^{;\alpha}\Box\omega
    -16\omega_{;\alpha\beta}\omega^{;\alpha}\omega^{;\beta}\}. \\
\end{array}
\end{equation}
The relation (\ref{b1}) acts as a condition on the dynamically
allowed configuration of conformal factor $\omega$. It selects the
physical metric as a metric in which the trace anomaly is replaced
by the standard conformal symmetry breaking term $\mu^2\psi^2$. The
physical characteristics of this metric depend essentially on the
boundary conditions imposed on the function $\psi$ and $\omega$. In
the subsequent two sections we study two examples of boundary
conditions with different physical characteristics.

\section{Asymptotic vacuum condition}

In this section, we investigate the behavior of the conformal frame
under the assumption that the background metric is approximated by
an asymptotically flat metric.

We first assume that the field $\omega$ is constant at
sufficiently large space-like distances, namely
\begin{equation}
\omega(x\rightarrow i_0)=const. \label{20}
\end{equation}
where $i_0$ denotes the space-like infinity. This condition means
that there is no distinction between the physical metric and the
background metric at space-like infinity. In other words we ignore
the back-reaction effect at space-like infinity. If a boundary
condition of this type is applied we get from the asymptotic
relation (\ref{20})
\begin{equation}
\Sigma_{\alpha\beta}(x\rightarrow i_0)=
3\mu^2\psi^{2}\bar{g}_{\alpha\beta}- \tau_{\alpha\beta}
 \label{21-a},\end{equation}
which follows directly from the gravitational equation (\ref{a5}).
By choosing $\mu=0$ at space-like infinity, the contribution to the
quantum stress tensor
 comes from the variation of $\psi$.
If we assume that the conformal frame corresponds asymptotically to
the Einstein frame at space-like infinity (constant configuration of
$\psi$ at $i_0$) we get from (\ref{21-a})
\begin{equation}
\Sigma_{\alpha\beta}(x\rightarrow i_0)\rightarrow 0.
\label{21-b}\end{equation}

This relation shows that for an asymptotically flat background
metric a Hadamard state which characterized by
$\Sigma_{\alpha\beta}$ looks like the vacuum of the Minkowski space
at sufficiently large space-like distances. Therefore the
super-selection rule (\ref{a4}) under the boundary conditions used
in this section may be considered as the curved space analogue of
DHR-super-selection rule \cite{8} in flat space quantum field
theory. In this case the whole quantum stress tensor vanishes at
space-like infinity. This kind of behavior is desirable only in
theories for which there is no asymptotic flow of vacuum energy at
space-like infinity (no particle creation). In such theories the
asymptotic flat background metric should reasonably taken as
globally static. For problems concerning a non-vanishing vacuum
energy at infinity, specifically  a non-vanishing asymptotic
radiation, other boundary conditions should be applied.

\section{Asymptotic particle creation}

In this section we consider the Schwarzschild metric as a background
metric
\begin{equation}
ds^2=(1-\frac{2GM}{r})dt^2-(1-\frac{2GM}{r})^{-1}dr^2-r^2d\theta^2-r^2\sin^2\theta
d\phi^2.
\end{equation}
here $G$ is the gravitational coupling and $M$ is the mass of the
black hole. We proceed to study the physical characteristics of the
physical metric $\bar{g}_{\alpha\beta}$ in (\ref{r}) with the
conformal factor solutions of equation (\ref{b1}). We show that the
corresponding conformal factor describes an outward flux of
radiation. We derive this result under the assumption $\mu=0$.
Equations (\ref{11}) and (\ref{b1}) in the Schwarzschild background
metric lead to
\begin{equation}\label{b3}
       \Box\psi-\Box\omega\psi+\omega_{;\alpha}\omega^{;\alpha}\psi-2\psi_{;\alpha}\omega^{;\alpha}=0
\end{equation}
\begin{equation}\label{b4}
      \frac{M^2}{15r^6}+6\Box(\Box\omega)+8(\Box\omega)^2
    -8\omega_{;\alpha\beta}\omega^{;\alpha\beta} -8\omega_{;\alpha}\omega^{;\alpha}\Box\omega
    -16\omega_{;\alpha\beta}\omega^{;\alpha}\omega^{;\beta}=0
     \end{equation}
We focus ourselves on those solutions having a non-vanishing flux of
energy momentum at infinity. For this purpose we take the solutions
satisfying these asymptotic conditions
\begin{equation}\label{b5}
\omega(t,r)=\frac{U(t-r)}{r^2}+{\mathcal{O}}(\frac{1}{r^3})\hspace{0.5cm},
\hspace{0.5cm}\lim_{r\rightarrow\infty}\psi=\psi_{0}=const.
\end{equation}
here $U$ is a smooth bounded arbitrary function of the retarded
time. Clearly these solutions imply a non-vanishing energy momentum
tensor at space-like infinity through the equation (\ref{a5}),
namely
\begin{equation}\label{b10}
 \Sigma_\alpha^\beta(r\rightarrow \infty)\propto\psi^2_0\{\frac{\ddot{U}}{r^2}\left(%
\begin{array}{cccc}
  -1 & 1 & 0 & 0\\
  -1 & 1 & 0 & 0 \\
  0 & 0 & 0 & 0 \\
  0 & 0 & 0 & 0 \\
\end{array}%
\right)+{\mathcal{O}}(\frac{1}{r^3})\}
\end{equation}
here the over-dot indicates differentiation with respect to $t-r$.
Thus at space-like infinity the contribution to the quantum stress
tensor
 comes from the variation of $\omega$. In other words, the effect of back-reaction can provide
 a non-vanishing asymptotic radiation.

 We select a state by the condition $\psi_0^2\ddot{U}\propto\frac{1}{G^2M^2}$. Under this assumption one
 can determine the
 power of this radiation (luminosity) via equation (\ref{b10}). At the first order the result is
\begin{equation}\label{b6}
P\propto\frac{1}{G^2M^2}.
\end{equation}
We infer that the amount of energy carried away by this radiation is
the same as the energy loss by the black hole. Therefore, the power
$P$ of this radiation is the rate of loss of total energy of the
black hole, that is
\begin{equation}\label{b7}
P=-\frac{dM}{dt}.
\end{equation}
Equating $P$ in equations (\ref{b6}) and (\ref{b7}) gives
\begin{equation}
-\frac{dM}{dt}\propto\frac{1}{G^2M^2}.
\end{equation}
which describes the black hole evaporation.

\section{Summary}

In this paper, we proposed a method to assign a conserved quantum
stress tensor with an anomalous trace to the state-dependent part of
the two-point function of a scalar quantum field conformally coupled
to a gravitational background. To characterize the local Hadamard
states, a super-selection rule was applied in form of a
self-consistent dynamical model which describes the gravitational
coupling of the quantum stress tensor to a dynamical metric. The
consequence of this super-selection rule was studied under two
different boundary conditions.

These considerations provided a vanishing vacuum energy at
space-like infinity in the case of ignoring the back-reaction effect
but a non-vanishing vacuum energy in the presence of this effect.


\begin{thebibliography}{99}
\bibitem{1} Hadamard J, Lectures on Cauchy's Problem in Linear Partial Diffrential
 Equations, Yale University Press, New Haven, (1923)
\bibitem{2} Adler S, Lieberman J and Ng Y. J, Ann. Phys. 106, 279 (1977)
\bibitem{3} Salehi H, Bisabr Y and Ghafarnejad H, J. Math. Phys. 41,
4582 (2000)
\bibitem{4}Brown M. R, J. Math. Phys. 25, 136 (1984)
\bibitem{5}Wald R. M, Phys. Rev. D. V17, N6 (1978)
\bibitem{6}Wald R. M, Ann. Phys. 110, 472 (1978)
\bibitem{7}  Birrell N. D and Davies P. C. W, Quantum fields in Curved
Space, Cambridge University Press, (1982).
\bibitem{7a} J.L. Synge, Relativity: The General Theory (North Holland, Amsterdam, 1966).

\bibitem{8}Hagg R, Local Quantum Physics, Springer (1992)


\end{thebibliography}
\end{document}